\documentclass[%
 reprint,
showpacs,preprintnumbers,
 amsmath,amssymb,
 aps,
prc,
superscriptaddress]{revtex4-1}

\usepackage{graphicx}
\usepackage{dcolumn}
\usepackage{bm}

\begin{document}

\preprint{APS/123-QED}

\title{Hyperonic neutron stars: reconciliation between nuclear properties and NICER and LIGO/VIRGO results  }

\author{Luiz L. Lopes}
\email{llopes@cefetmg.br}

\affiliation{%
 Centro Federal de Educa\c{c}\~ao Tecnol\'ogica de Minas Gerais Campus VIII; CEP 37.022-560, Varginha - MG - Brasil
}%

\date{\today}

\begin{abstract}
 Using an extended version of the Quantum Hadrodynamics (QHD), I propose a new microscopic
equation of state (EoS) able to correctly reproduce the main properties of symmetric nuclear matter
at the saturation density, as well as produce massive neutron stars and satisfactory results for the
radius and the tidal parameter {$\Lambda$}. I show that this EoS can
reproduce at least 2.00 solar masses neutron star even when hyperons are present. The constraints about the radius of a 2.00$M_\odot$ and the minimum mass that enables direct Urca effect are also checked. 
 
\end{abstract}


\maketitle

\section{Introduction}

Our knowledge of  nuclear physics and nuclear astrophysics took a great leap in the last decade. From nuclear masses analyze~\cite{Wang},  passing through nuclear resonances~\cite{Colo,Reinhard}, and heavy ion collisions (HIC)~\cite{Pagano}; we are able to constraint six parameters of the symmetric nuclear matter at the saturation point: the saturation density itself ($n_0$), the effective nucleon mass ($M_N^*/M_N$), the incompressibility ($K$), the symmetry energy ($S_0$) and its slope ($L$), as well  the binding energy per
baryon ($B/A$)~\cite{Glen}.

On the other hand, quiescent analyzes~\cite{Heinke} and Shapiro delay measurements~\cite{Antoniadis}  constraint the radius of the canonical star, as well the minimum mass that a microscopic EoS needs to reproduce, in order to be considered valid. The constraints improved in the last years with the results coming for the NICER x-ray telescope~\cite{NICER1}, and also, from the analyzes of the so-called GW170817 event, detected by the  LIGO/VIRGO gravitational wave observatories~\cite{GW17}. 
Moreover, the newly opened gravitational waves window, give us an additional constraint: the dimensionless tidal deformation parameter $\Lambda$.
Another recent constraint is related to how squeezable are the neutron stars. In the past, it was believed that very massive neutron stars should
be smaller than less massive ones. However, a very recent study~\cite{NICER3}
shows that a two solar masses and a canonical mass of 1.4 $M_\odot$
have very similar radii. This fact puts an additional constraint on the microscopic EoS.

Nowadays one of the open issues in nuclear astrophysics is the content of the inner core of massive neutron stars. Because of the Pauli principle, as the number density increases, the  Fermi energy of the nucleons exceeds the mass of heavier baryons, and the conversion of some nucleons into hyperons becomes energetically favorable. Among others, two extensive studies about the hyperon threshold~\cite{hyp2010,lopesnpa} show that hyperons are  ultimately inevitable. Otherwise, it's been well known for a long time that the hyperon onset softens the EoS. In some cases, this softening of the EoS puts the maximum mass below the observational limits of massive neutron stars. This possible conflict between theory and observation is called hyperon puzzle.

The hyperon puzzle can be avoided assuming a very stiff EoS, as GM1~\cite{Glen2} or NL3~\cite{Lala}, which produce very massive neutron stars, despite the hyperon threshold. However, due to the refinement of the constraint in both, nuclear properties, as well in astrophysical observations, these models cannot be faced as realistic anymore. { 
Another possibility is the use of strongly repulsive three-body forces, as shown in ref.~\cite{nomorehyperons}. As the strength of the three-body forces is not yet constrained, this is an open issue.}

In this work, I present a new microscopic EoS via an extended version of the QHD Lagrangian, where besides the traditional non-linear $\sigma\omega\rho$ mesons, also employs the strangeness hidden $\phi$ vector meson, which couples only to the hyperons and a non-linear $\omega-\rho$ coupling, present in models like the IUFSU~\cite{IUFSU}. 
Also, to fix the hyperon-meson coupling constant, I use complete symmetry arguments as presented in ref.~\cite{Lopes2013}.

The formalism, the parametrization, and the constraints of symmetric nuclear matter at the saturation point, and supra-nuclear densities are presented in section II; the astrophysical results and the astrophysical constraints are presented in section III; finally,
the conclusions are drawn in section IV.

\section{Formalism \label{sec2}}

The non-linear $\sigma\omega\rho$ QHD has the following Lagrangian density~\cite{Glen}:

\begin{eqnarray}
\mathcal{L}_{QHD} = \sum_B \bar{\psi}_B[\gamma^\mu(\mbox{i}\partial_\mu  - g_{B\omega}\omega_\mu   - g_{B\rho} \frac{1}{2}\vec{\tau} \cdot \vec{\rho}_\mu)+ \nonumber \\
- (M_B - g_{B\sigma}\sigma)]\psi_B  -U(\sigma) +   \nonumber   \\
  + \frac{1}{2}(\partial_\mu \sigma \partial^\mu \sigma - m_s^2\sigma^2) - \frac{1}{4}\Omega^{\mu \nu}\Omega_{\mu \nu} + \frac{1}{2} m_v^2 \omega_\mu \omega^\mu+  \nonumber \\
 + \frac{1}{2} m_\rho^2 \vec{\rho}_\mu \cdot \vec{\rho}^{ \; \mu} - \frac{1}{4}\bf{P}^{\mu \nu} \cdot \bf{P}_{\mu \nu}  , \label{s1} 
\end{eqnarray}
in natural units. 
 The sum in $B$ can run only over the nucleons, or the entire baryon octet; $\psi_B$  is the baryonic  Dirac field.  The $\sigma$, $\omega_\mu$ and $\vec{\rho}_\mu$ are the mesonic fields.
 The $g's$ are the Yukawa coupling constants that simulate the strong interaction, $M_B$ is the baryon mass,  $m_s$, $m_v$,  and $m_\rho$ are
 the masses of the $\sigma$, $\omega$, and $\rho$ mesons respectively.
 The antisymmetric mesonic field strength tensors are given by their usual expressions as presented in~\cite{Glen}.
  The $U(\sigma)$ is the self-interaction term introduced in ref.~\cite{Boguta} to fix the incompressibility, given by:
 
 \begin{equation}
U(\sigma) =  \frac{\kappa M_N(g_{\sigma} \sigma)^3}{3} + \frac{\lambda(g_{\sigma}\sigma)^4}{4} \label{s2} ,
\end{equation}

\noindent and , $\vec{\tau}$ are the Pauli matrices.
Now, besides the traditional non-linear $\sigma\omega\rho$ QHD, it is needed that we introduce two additional terms.  The first one is the strangeness hidden $\phi$ vector meson, which couples only with the hyperons, not affecting the properties of symmetric matter:

\begin{equation}
\mathcal{L}_\phi = g_{Y,\phi}\bar{\psi}_Y(\gamma^\mu\phi_\mu)\psi_Y + \frac{1}{2}m_\phi^2\phi_\mu\phi^\mu - \frac{1}{4}\Phi^{\mu\nu}\Phi_{\mu\nu} , \label{EL3} ,
\end{equation}

\noindent as pointed in ref.~\cite{lopesnpa,Rafa2005,Lopes2020a,Weiss1}, this vector channel is crucial to obtain massive hyperonic neutron stars. The second one is a non-linear  $\omega$-$\rho$ coupling meson as present in the IUFSU model~\cite{IUFSU}:

\begin{equation}
 \mathcal{L}_{\omega\rho} = \Lambda_{\omega\rho}(g_{\rho}^2 \vec{\rho^\mu} \cdot \vec{\rho_\mu}) (g_{\omega}^2 \omega^\mu \omega_\mu) ,
\end{equation}

\noindent which is necessary to correct the slope of the symmetry energy ($L$) and has a strong influence on the radii and tidal deformation of the neutron stars~\cite{Rafa2011,dex19jpg}. 

To produce beta stable matter, with zero net charge, I also add leptons as a free Fermi gas. The detailed calculation of the EoS for symmetric nuclear matter, as well for beta stable matter in the QHD formalism is out of the scope of this work, but is well documented and can easily be found in the literature~\cite{Glen,Serot,Menezes2021s}. In the same sense, the calculation of the six nuclear parameters at the saturation density ($n_0,~ M_N^*/M_N,~K,~S_0,~L,~B/A$) can be found in ref.~\cite{Glen,Rafa2011} and the references therein.

\subsection{Parametrization and nuclear constraints} 

The parametrization utilized in this work, as well the predictions of this model for the symmetric nuclear matter are presented in Tab.~\ref{TL1}. The nuclear constraints  at the saturation density are also in Tab.~\ref{TL1} and are taken  from two extensive review articles, ref.~\cite{Dutra2014,Micaela2017}. Besides, the masses of the particles are the physical ones. The meson masses are $m_\omega$ = 783 MeV, $m_\rho$ = 770 MeV, $m_\phi$ = 1020 MeV, $m_\sigma$ = 512 MeV, the baryon octed masses are $M_N$ = 939 MeV, $M_\Lambda$ = 1116 MeV, $M_\Sigma$ = 1193 MeV, $M_\Xi$ = 1318 MeV, and the lepton masses are $m_e$ = 0.51 MeV, $m_\mu$ = 105.6 MeV.

\begin{widetext}
\begin{center}
\begin{table}[ht]
\begin{center}
\begin{tabular}{|c|c||c|c|c||c|}
\hline 
  & Parameters & &  Constraints  & This model  \\
 \hline
 $(g_{N\sigma}/m_s)^2$ & 12.108 $fm^2$ &$n_0$ ($fm^{-3}$) & 0.148 - 0.170 & 0.156 \\
 \hline
  $(g_{N\omega}/m_v)^2$ & 7.132  $fm^2$ & $M^{*}/M$ & 0.6 - 0.8 & 0.69  \\
  \hline
  $(g_{N\rho}/m_\rho)^2$ & 4.801  $fm^2$ & $K$ (MeV)& 220 - 260                                          &  256  \\
 \hline
$\kappa$ & 0.004138 & $S_0$ (MeV) & 28.6 - 34.4 &  31.2  \\
\hline
$\lambda$ &  -0.00390 & $L$ (MeV) & 36 - 86.8 & 74\\
\hline 
$\Lambda_{\omega\rho}$ &  0.0185 & $B/A$ (MeV) & 15.8 - 16.5  & 16.2  \\ 
\hline
- &  - & $S(2n_0)$ (MeV) & 38 - 64  & 52.4  \\ 
\hline
- &  - & $p(2n_0)$ (MeV/fm$^3$) & 11.2 - 38.7  & 16.4  \\ 
\hline
\end{tabular}
 
\caption{ Parameters of the model utilized in this work and their prediction for the symmetric nuclear matter  at the saturation density; the phenomenological constraints are taken from ref.~\cite{Dutra2014,Micaela2017,s3,tidal1}, as well the recent PREX2 results~\cite{PREX2}. } 
\label{TL1}
\end{center}
\end{table}
\end{center}
\end{widetext}

\begin{figure}[h] 
\begin{centering}
 \includegraphics[angle=270,
width=0.5\textwidth]{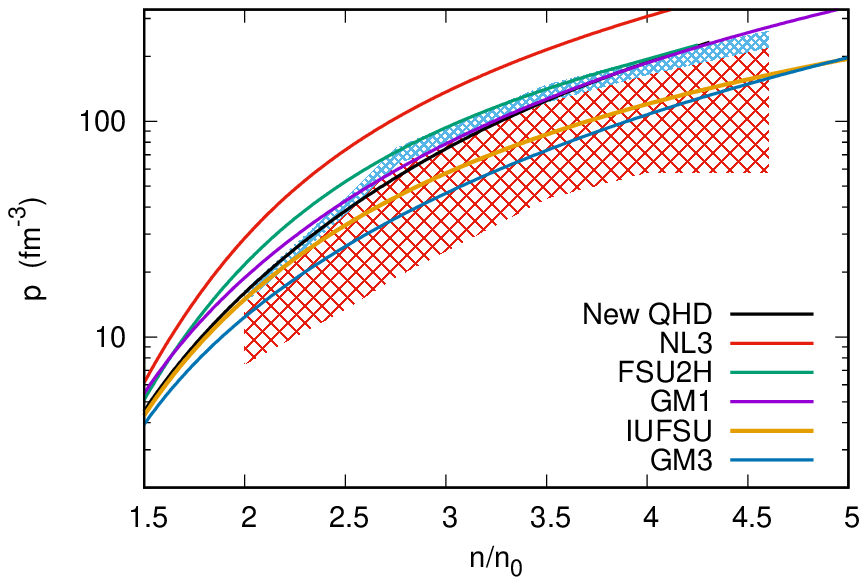}
\caption{(Color online) The pressure as a  function of the density;  the original constraint from ref.~\cite{Daniel} (in red), and its value extended by 20$\%$ (in blue), to reconcile with the constraint presented in ref.~\cite{Steiner2013}. } \label{F1}
\end{centering}
\end{figure}

As displayed, this new parametrization is able to fulfill all the constraints at the saturation density. Notice also that the new PREX2 results point that the slope is constrained between 106 $\pm$ 37 MeV~\cite{PREX2}. Combined with ref.~\cite{Micaela2017}, the slope now lies between 69 and 86.8 MeV. The value $L =  74$ MeV fulfill it. As PREX2 still needs confirmation, I only point it as a curiosity.  Besides the saturation point, there are also a couple of tries in order to constraint the pressure of symmetric nuclear matter at supranuclear densities. 
In ref.~\cite{Daniel}, the pressure from 2 to 4.6 times the saturation density was constrained by HIC analyzes. On the other hand, from transiently accreting neutron stars in quiescence, ref.~\cite{Steiner2013} constraint the pressure for densities up to 1.0 $fm^{-3}$. The main problem is that for high densities, the region constrained in ref.~\cite{Steiner2013} is broader than the region constrained in ref.~\cite{Daniel}.
To overcome this issue, ref.~\cite{Dutra2014} assumes the region from ref.~\cite{Daniel} plus an increase of 20$\%$. Here, I follow this prescription and plot in Fig.~\ref{F1} the density-dependent pressure of the presented model altogether with some well-known relativistic parametrizations. The hatched area in red is the constraint originally presented in ref.~\cite{Daniel}. The increase by 20$\%$ is the hatched area in blue.

As can be seen, the model is not flawless. If we assume the original constraint in red, the EoS discussed in this work have an over-pressure, and only fulfill this constraint in the range between 2.52 to 3.43 times the saturation density, which corresponds to 35$\%$ of the interval (2 - 4.6) $n_0$. But, if we assume that the upper limit can be increased by 20$\%$, as done in ref.~\cite{Dutra2014}, therefore this EoS fulfills the constraint from 2 to 4.20 times the saturation density, which turns out to be 85$\%$ of the range. When compared with other parametrizations of the QHD, we see that the presented model presents a higher degree of agreement (85$\%$) in comparison with very stiff EoS, as NL3~\cite{Lala} (0$\%$), FSU2H~\cite{Tolos} (50$\%$), and GM1~\cite{Glen2} (68$\%$), but lower degree of agreement when compared with soft EoS as GM3~\cite{Glen2} and IUFSU~\cite{IUFSU} (both 100$\%$).

Another two recent constraints at supra-nuclear densities are the symmetry energy at two times the saturation density, $S(2n_0)$, and the pressure of the symmetric matter at the same point $p(2n_0)$. Ref.~\cite{s3}  bounds $S(2n_0)$ in the range between  51 $\pm$ 13 MeV at 68$\%$ confidence level, while ref.~\cite{tidal1} fixed $p(2n_0)$ between 11.2 MeV/fm$^{3}$ and 38.7 Mev/fm$^{3}$ at 90$\%$ level. These values are also presented in Tab.~\ref{TL1}, and the new parametrization also agrees with both.

\section{Astrophysical results and constraints \label{sec3}}

Now I turn my attention to construct a beta stable, charge-neutral EoS, with and without hyperons. When hyperons are present, it is crucial to define the strength of the hyperon-mesons coupling constants. The only well-known parameter is the $\Lambda^0$ hyperon potential depth, $U_\Lambda$ = -28 MeV. We can find in the literature several approaches to fix the hyperon-mesons coupling constants, from universal couplings~\cite{Glen2}, trough to fixed potential depths~\cite{lopesnpa,Weiss2}, non-fixed potential depths~\cite{Weiss1,Lopes2013,Stone2021,Tsu}, with and without symmetry group arguments~\cite{Rafa2011,Biswal} and so  on.  Here, I follow ref.~\cite{Lopes2013} and assume that both, vector and scalar mesons are constrained by SU(3) symmetry group~\cite{Tsu,Swart}. In this case, the relative strength of the hyperon-mesons coupling constants are:

\begin{equation}
 \frac{g_{\Lambda,\omega}}{g_{N,\omega}} = \frac{4 + 2\alpha_v}{5 + 4 \alpha_v}; \quad
 \frac{g_{\Sigma,\omega}}{g_{N,\omega}} = \frac{8 - 2\alpha_v}{5 + 4 \alpha_v}; \quad
  \frac{g_{\Xi,\omega}}{g_{N,\omega}} = \frac{5 - 2\alpha_v}{5 + 4 \alpha_v}; \label{eomega}
\end{equation}
\noindent for the $\omega$ meson,

\begin{eqnarray}
 \frac{g_{\Lambda,\phi}}{g_{N,\omega}} = \sqrt{2} \bigg (\frac{ 2\alpha_v - 5}{5 + 4 \alpha_v} \bigg ); \quad
 \frac{g_{\Sigma,\phi}}{g_{N,\omega}} = \sqrt{2} \bigg (\frac{ - 2\alpha_v - 1}{5 + 4 \alpha_v} \bigg ); \quad \nonumber \\
  \frac{g_{\Xi,\phi}}{g_{N,\omega}} = \sqrt{2} \bigg (\frac{ - 2\alpha_v - 4}{5 + 4 \alpha_v} \bigg ) ;\nonumber \\ \label{ephi}
\end{eqnarray}
\noindent for the $\phi$ meson,

\begin{equation}
 \frac{g_{\Lambda,\rho}}{g_{N,\rho}} = 0; \quad
 \frac{g_{\Sigma,\rho}}{g_{N,\rho}} =  2\alpha_v; \quad
  \frac{g_{\Xi,\rho}}{g_{N,\rho}} = -(1 - 2\alpha_v); \label{erho}
\end{equation}
\noindent for the $\rho$ meson, and finally:
\begin{eqnarray}
 \frac{g_{\Lambda,\sigma}}{g_{N,\sigma}} =  \frac{ 10 + 6\alpha_s}{13 + 12 \alpha_s} ; \quad
 \frac{g_{\Sigma,\sigma}}{g_{N,\sigma}} = \frac{ 22 - 6\alpha_s}{13 + 12 \alpha_s}; \quad \nonumber \\
  \frac{g_{\Xi,\sigma}}{g_{N,\sigma}} = \frac{ 13 - 6\alpha_s}{13 + 12 \alpha_s}; \nonumber \\  \label{esigma}
\end{eqnarray}
\noindent for the $\sigma$ meson. When we set $\alpha_v$ = 1, we recover the SU(6) parametrization for the vector mesons~\cite{Pais}.  Now, to constraint $\alpha_v$ to $\alpha_s$ we use the well-known $\Lambda^0$ potential depth. Therefore, for a given value of $\alpha_v$, we determine $\alpha_s$ to obtain $U_\Lambda$ = -28 MeV. This approach allows us to fix every hyperon-meson coupling constant using only one free parameter: $\alpha_v$. Also, it is worth to point that the results from eq.~\ref{eomega} to eq.~\ref{esigma} are fully model-independent. The results obtained are presented in Tab.~\ref{T2}.
{ The  $U_\Sigma$ and $U_\Xi$ potential depths are also calculated. As can be see the values for $\alpha_v$ equal to 1.00 and 0.75 are more repulsive than the values recently found in ALICE femtoscopic measurements ($U_\Sigma$ = +15 MeV, $U_\Xi$ = -4 MeV)~\cite{ALICE}, but still close to what is currently accepted in the literature ($\pm$ 40 MeV)~\cite{Weiss1}. An alternative would be to use SU(3) symmetry group to fix the vector mesons coupling constants while use the potential depth to fix the scalar meson coupling constant as made in ref.~\cite{lopesnpa,Weiss2,Rather}. However, in this approach, we would  need three free parameters instead of just one.}

\begin{table}[ht]
\begin{center}
\begin{tabular}{|c||c|c|c|}
\hline
 $\alpha_v$& 1.00 & 0.75 & 0.50 \\
\hline
$\alpha_s$   & 1.582 & 1.240 & 0.911\\
 \hline
 $U_\Sigma$  (MeV)  & 34 & 29 & 46\\
 \hline
 $U_\Xi$ (MeV)   & 41 & 39 & 52\\
 \hline
 \end{tabular} 
\caption{Relation of $\alpha_v$ with $\alpha_s$ in order to produce $U_\Lambda$ = -28 MeV, { as well the predicted values of $U_\Sigma$ and $U_\Xi$.} When $\alpha_v$ = 1.00 we recover the SU(6) symmetry group.} 
\label{T2}
\end{center}
\end{table}

\begin{figure*}[ht]
\begin{tabular}{cc}
\includegraphics[width=0.33\textwidth,angle=270]{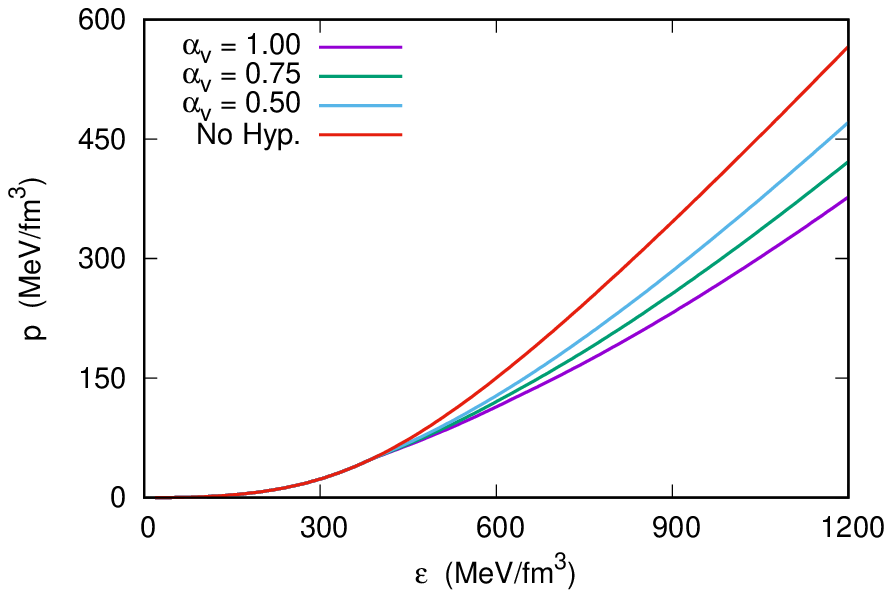} &
\includegraphics[width=0.33\textwidth,,angle=270]{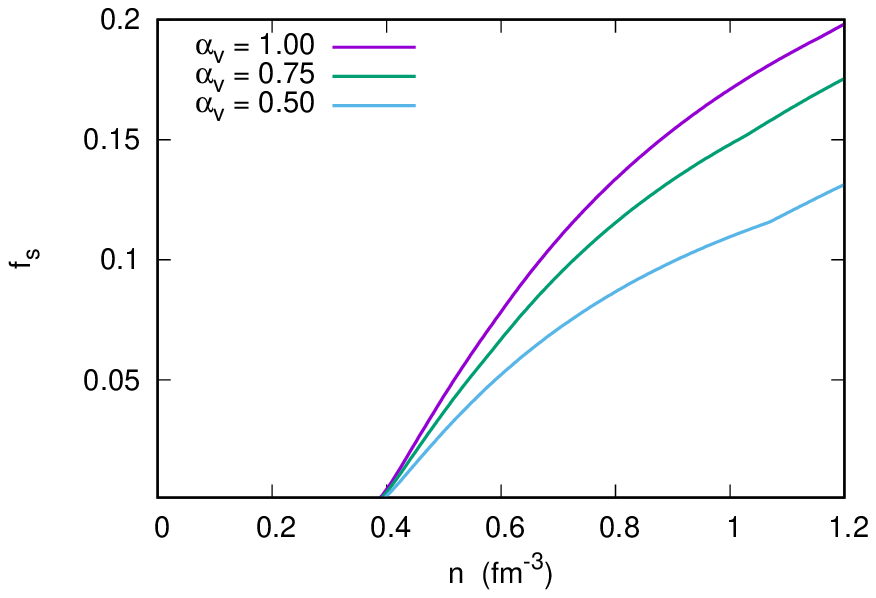} \\
\end{tabular}
\caption{(Color online) EoS (left) and the strangeness fraction $f_s$ (right) for different values of $\alpha_v$. Higher the value of $\alpha_v$, higher is the value of $f_s$ and softer is the EoS.} \label{F2}
\end{figure*}

\begin{figure*}[ht]
\begin{tabular}{cc}
\includegraphics[width=0.33\textwidth,angle=270]{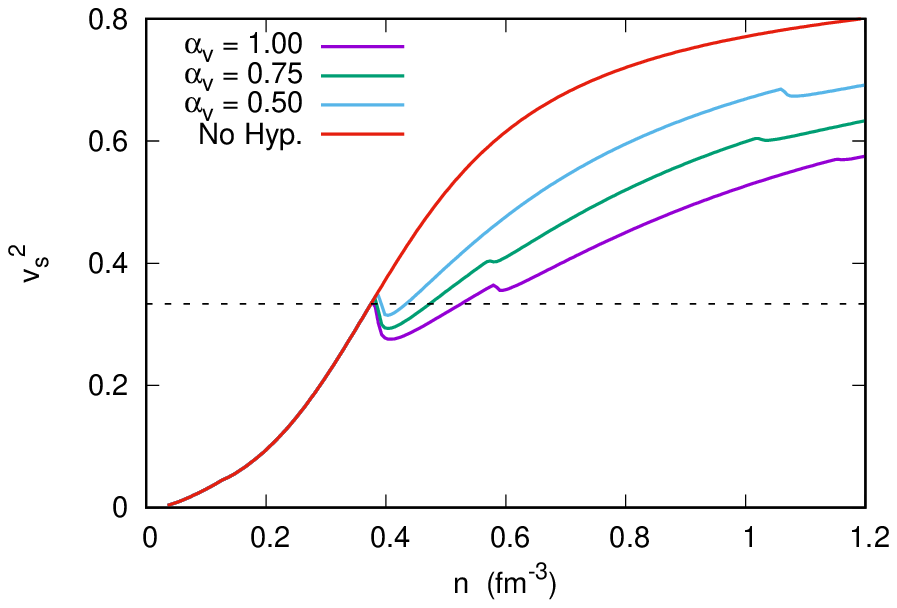} &
\includegraphics[width=0.33\textwidth,,angle=270]{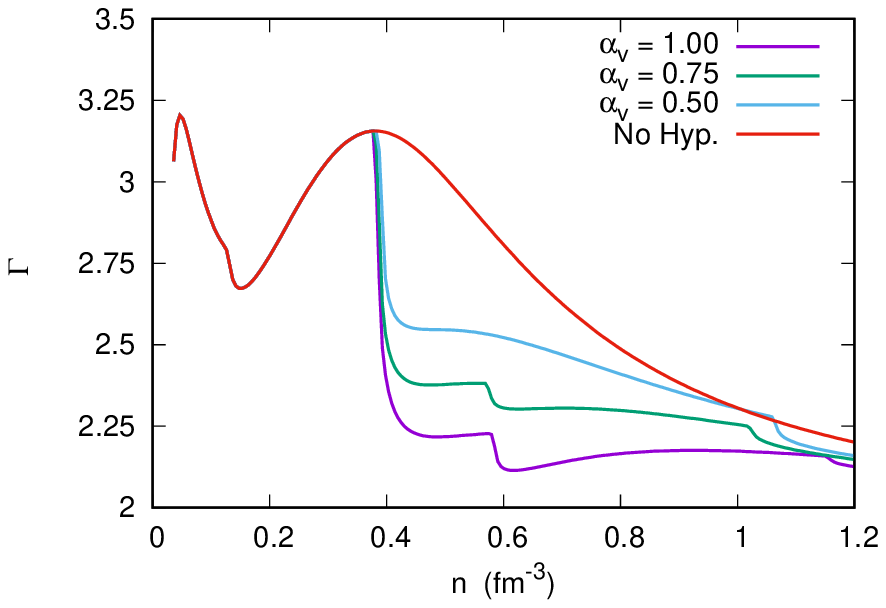} \\
\end{tabular}
\caption{(Color online) Speed of sound  (left) and the adiabatic index $\Gamma$ (right) for different values of $\alpha_v$. The horizontal dotted line is the conformal limit: $v_s^2 = 1/3$.} \label{F3}
\end{figure*}

In fig.~\ref{F2}, I plot the EoS for beta stable, charge-neutral matter with and without hyperons for different values of $\alpha_v$.
I also plot the strangeness fraction $f_s$ for each value of $\alpha_v$,  once previous works showed that its value is strongly related to the maximum mass~\cite{Weiss2,lopesnpa}. The strangeness fraction is defined as: ($f_s = \sum (n_i|s_i|)/{3n}$)~\cite{Rafa2011}. It is also worth pointing that the sequence of hyperon threshold for each value of $\alpha_v$ is always the same: the first one is the $\Lambda^0$, followed by the $\Sigma^-$ and the $\Xi^-$, except when $\alpha_v$=0.50, where the $\Sigma^{-}$ is absent.

As expected, there is an inversely proportional relationship between the stiffness of the EoS with the
strangeness fraction at high densities. In turn, the strangeness content of the EoS is directly linked to the value of $\alpha_v$. Higher the value of $\alpha_v$, higher is the value of $f_s$. Also, despite the value of $\alpha_v$, hyperons are always present, corroborating the results present in ref.~\cite{hyp2010,lopesnpa}.

{ The effects of the onset of the strangeness content particles can be better seen by analyzing the square of the speed of sound, $v_s^2$ as well the adiabatic index, $\Gamma$. They are defined as:}

\begin{equation}
 v_s^2 = \bigg | \frac{\mbox{d}p}{\mbox{d}\epsilon} \bigg | \quad \mbox{and} \quad  \Gamma = \frac{(p +\epsilon)}{p}\frac{\mbox{d}p}{\mbox{d}\epsilon} ,
\end{equation}
{ \noindent and the results are displayed in fig.~\ref{F3}. As can be seen, the onset of the hyperons causes the presence of kinks and peaks in the speed of sound. Therefore, the onset of the $\Lambda^0$ hyperon around 0.4 fm$^{-3}$ can  easily  be identified. A very important value is $v_s^2 = 1/3$, called conformal limit~\cite{Stone2021}. In the limit of very high densities $n > 40n_0$, pQCD indicates that the $v^2_s$ should approach 1/3 from below~\cite{PRLTan,Kojo}. Nevertheless, in the range of interest for neutron star interiors, $n < 8n_0$, we see that the conformal limit is always violated in the proposed parametrization, while the causal limit, $v_s^2 < 1$, is always respected.

Another important physical quantity is the adiabatic index, $\Gamma$,
a sensitive indicator of phase changes in stellar matter and the stability with respect to vibrations and pulsation  of the star~\cite{Haensel2002,Haensel2008}. For multicomponent matter,
exhibit jumps at densities coincident with density thresholds of
individual components, signaling phase transitions and/or changes
in the make-up of the matter~\cite{Stone2021}. As showed in fig.~\ref{F3}, the $\Lambda^0$ threshold is strongly evidenced, causing a huge drop in the value of $\Gamma$. Smaller, but distinguishable peak around 0.6 $fm^{-3}$ points to the onset of $\Sigma^-$ hyperon for $\alpha_v$ = 1.00 and $\alpha_v$ = 0.75. As pointed out earlier, for $\alpha_v$ = 0.50, the $\Sigma^-$ is suppressed.  }

Concerning the neutron star properties and the astrophysical constraint, one of the most robust constraints is the minimum mass that an EoS must reproduce due to observational measures of massive stars via Shapiro delay.
Not long ago, the neutron star king was the PSR J0348+0432, with a mass of 2.01 $\pm 0.04$ $M_\odot$~\cite{Antoniadis}. The old king was deposed, and now the PSR J0740+6620 with a mass of approximately of 2.07 $\pm$ 0.07 $M_\odot$~\cite{NICER3}
rules over the sky. Notwithstanding, both pulsars send the same message: the EoS must be stiff enough to produce a two solar masses neutron star.
On the other hand, one of the most controversial constrain is the radius of the canonical 1.4$M_\odot$. Just in the last couple of years, different studies point to different values for their maximum allowed value. For instance,
a maximum value of 13.8 km, 14.2 km, 13.5 km, 12.6 km, 13.5 km, 12.8 km, 13.0 km, and 13.2 km was found
respectively in ref.~\cite{NICER1,NICER2,rad1,rad2,rad3,rad4,rad5,rad6}.
The average value of these studies is 13.32 km, and I use it as a constraint.

Also, as pointed in ref.~\cite{urcac}, any acceptable EoS shall not allow the direct Urca process to occur in NS with masses below 1.5$M_\odot$. The trigger to nucleonic direct Urca channel is directly related to the leptonic fraction $x_{DU}$.~\cite{Rafa2011,urcac}. On the other hand, when hyperons are present the situation is more complex. Hyperons disfavors the DU process due to the deleptonization of the star, and the nucleonic DU process may be forbidden.
However, channels for neutrino emission involving the hyperons are also possible but with lower efficiency. As in this work, the $\Lambda^0$ is always the first
and the most populous hyperon, I only consider its process. The nucleonic, $x_{DU}=Y_p$ and hyperonic, $x_{\Lambda^0}=Y_{\Lambda}$ fraction to enable DU process are~\cite{urcae}:

\begin{equation}
 x_{DU} = \frac{1}{1 + (1+ x_e^{1/3})^3}, \quad \mbox{and}
 \quad x_{\Lambda^0} = 0.032, \label{du}
\end{equation}
\noindent where $x_e$ = $n_e/(n_e + n_\mu)$. Notwithstanding,
the efficiency of the hyperonic DU process is only $4\%$ of the nucleonic one~\cite{urcae}.

Another important constraint comes from the GW170817 event, detected by the  LIGO/VIRGO gravitational wave telescopes: the dimensionless tidal deformation parameter $\Lambda$.
The tidal deformability of a compact object is a single parameter that quantifies how easily the object is deformed when subjected to an external tidal field. Larger tidal deformability indicates that the object is easily deformable. On the opposite side, a compact object with a smaller tidal deformability parameter is smaller, more compact, and it is more difficult to deform. Its value is defined as: $\Lambda ={2k_2}/{3C^5}$, where  C = ($GM/R$) is the compactness. The parameter $k_2$ is called Love number and is related to the metric perturbation. A complete discussion about the Love number and its calculation is
both, very extensive and well documented in the literature. Therefore, it is out of the scope of this work. I refer the interested reader to see ~\cite{tidal1,tidal2,tidal3,tidal4} and the references therein. Ref.~\cite{tidal1}
constraint the dimensionless tidal parameter for the canonical mass, $\Lambda_{1.4}$ in the range of 70-580. Here, I use 580 as a superior limit.

\begin{figure*}[ht]
\begin{tabular}{cc}
\includegraphics[width=0.33\textwidth,angle=270]{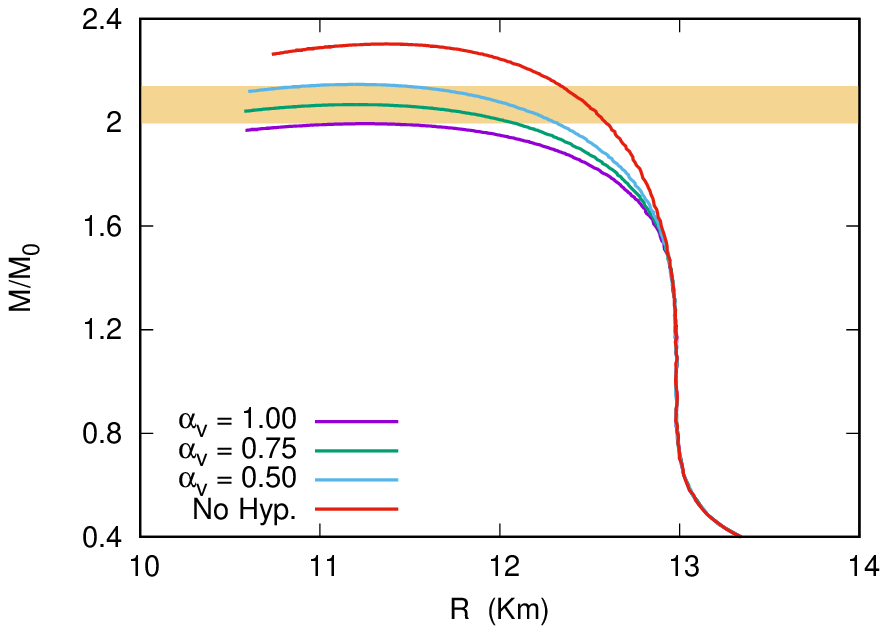} &
\includegraphics[width=0.33\textwidth,,angle=270]{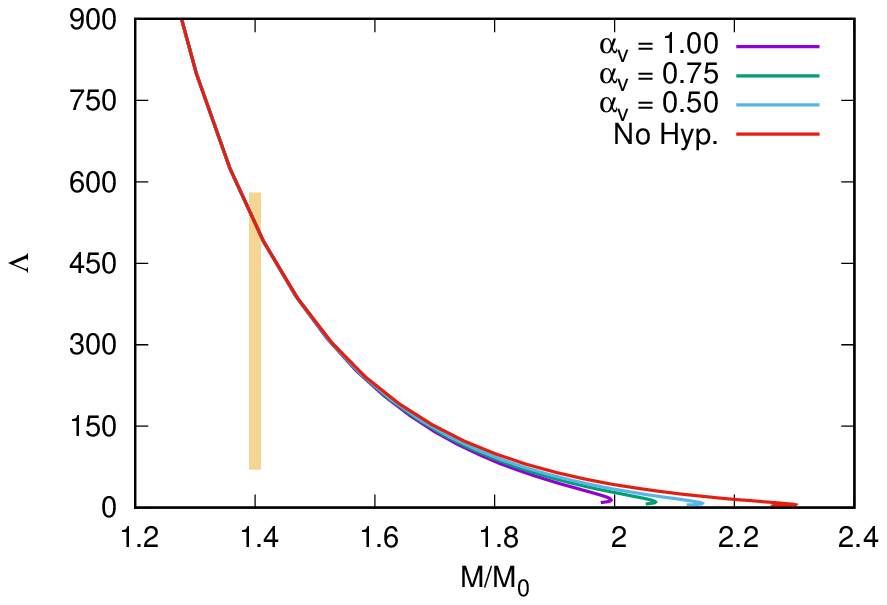} \\
\end{tabular}
\caption{(Color online) Mass radius relation (left) and the dimensionless tidal parameter $\Lambda$ (right). The hatched area on the left (right) is related to the uncertain about the mass of PSR J0740+6620 (uncertain about the value of $\Lambda_{1.4}$ in the GW170817 event~\cite{tidal1}).} \label{F4}
\end{figure*}

Finally, a very recent constraint comes from the measurement of the radius of the massive pulsar PSR J0740+6620  made by the NICER x-ray telescope. This is related to how squeezable are the neutron stars and allows us to put an inferior limit to the radius of a two solar mass object. A inferior bound of  11.4 km, 12.2 km, 11.6 km and 12.2 km was found respectively in ref.~\cite{NICER3,s1,s2,s3}. An average value of 11.85 km arises and I also use this value as a constraint to the EoS.

In fig.~\ref{F4}, I display both, the TOV equation solution~\cite{TOV}, as well the calculated value for the dimensionless tidal parameter $\Lambda$. The BPS EoS is used to modelling the neutron star crust~\cite{BPS}. The hatched areas are uncertain about the mass value of PSR J0740+6620~\cite{NICER3}, as well the uncertainty about the $\Lambda_{1.4}$~\cite{tidal1} value. The main neutron star properties and the astrophysical constraints are also displayed at Tab.~\ref{T3}.

\begin{table}[ht]
\begin{center}
\begin{tabular}{|c||c|c|c|c||c|}
\hline
 $\alpha_v$ &~1.00~&~0.75~&~0.50~& No hyp. & Constraint  \\
\hline
$M_{max}/M_\odot$   & 2.00 & 2.07 & 2.15 & 2.30 & $>$ 2.00   \\
 \hline
 $f_s$ ($M_{max}$) & 0.173 &  0.151 &  0.110 & 0.000 & - \\
 \hline
 $n_{max}$ (fm$^{-3}$) & 1.01 &  1.02 &  1.00 & 0.94 & - \\
 \hline
 $R_{1.4}$ (km) & 12.96  & 12.96  &  12.96 & 12.96 & $<$ 13.32\\
 \hline
  $\Lambda_{1.4}$& 527  & 527  & 527 & 527 & $<$ 580    \\
 \hline
$R_{2.0}$ (km) & 11.27  & 12.07  &  12.28 & 12.58  &  $>$ 11.85   \\
 \hline
 $f_s$ (2.0$M_\odot$) & 0.173 &  0.103 &  0.064 & 0.000 & - \\
 \hline
 $n_{DU}$ (fm$^{-3}$) & 0.42* &  0.43* & 0.43* & 0.47 & - \\
 \hline
 $M_{DU}/M_\odot$   & 1.50* & 1.52* & 1.55* & 1.73 & $>$ 1.50   \\
 \hline
 \end{tabular} 
\caption{Some neutron star properties and constraints for different values of $\alpha_v$. Results marked with * indicate hyperonic DU process, with only 4$\%$ of the efficiency of the nucleonic DU.} 
\label{T3}
\end{center}
\end{table}

As can be seen, even in the presence of hyperons, we are to produce a two solar mass neutron star. Also, as hyperons are not present at masses around 1.4$M_\odot$, the $R_{1.4}$ and $\Lambda_{1.4}$ always give the value of 12.96 km and 527 respectively. This then means that besides correctly
reproduce the six parameters of symmetric nuclear matter, this EoS also
fulfill the main astrophysical constraints: $M_{max} >2.00M_\odot$, $R_{1.4} < 13.32$,  $M_{DU} > 1.50M_\odot$, and $\Lambda_{1.4} < 580$. The only astrophysical constraint that is not satisfied for all models is the recent measure of the radius of the PSR J0740+6620,
as $\alpha_v$ = 1.00 produces a too low radius for a 2.00$M_\odot$.
Indeed, as can be seen in Tab.~\ref{T3}, there is a relation between the maximum possible mass and the radius of the 2.00$M_\odot$. Higher is the maximum mass, higher is the radius of 2.00$M_\odot$. This relation can be explained due to the different strangeness content in the core of the 2.00$M_\odot$ star. Higher the value of $f_s$, softer is the EoS, lower is the radius of the 2.00$M_\odot$ as well the maximum mass. Also, when hyperons are present, the nucleonic DU process is always forbidden and only the hyperonic DU process is present. Without hyperons, nucleonic DU happens for masses above 1.73$M_\odot$, which agrees with ref.~\cite{urcac}.

\section{Final Remarks}

In this work, I build a new EoS that can satisfy all six main constraints at the nuclear saturation point. This EoS also is able to fulfill the experimental constraint about the pressure at supra-nuclear densities for 85$\%$ of the interval presented in the combined ref.~\cite{Daniel,Steiner2013} and produce good values for $S(2n_0)$ and $p(2n_0)$, in agreement with ref.~\cite{s3,tidal1}.

In the realm of astrophysical constraint, this EoS is able to reproduce masses above 2.00$M_\odot$ even when hyperons are present. It also produces $R_{1.4}~<~13.32$, km $M_{DU} > 1.50M_\odot$,   $\Lambda_{1.4}~<~580$ and $R_{2.0}~>11.85$, km; thus satisfying all the main astrophysical bounds.

{ The applications of this new parametrization can be extended in a large branch of areas. As I presented only infinite  symmetric and beta stable nuclear  matters at zero temperature, results concerning finite temperature effects, finite nuclei, magnetic and electric fields, pasta phase, and so on, are still open.} 

Before I finish, it is worth pointing out that PREX2 experiments have pointed for much larger values for both, the slope $L$, as well the tidal parameter $\Lambda_{1.4}$ and the radius of the canonical star~\cite{PREX2}.
These results can be far beyond the most acceptable values coming from several different studies and still needs confirmation.
Nevertheless, if the PREX2 results turn out to be true, that does not imply a strong modification in the presented EoS. Only the $g_{\rho,N}$ coupling constant and the $\Lambda_{\omega\rho}$ parameter will be needed to be modified. The same is true if the slope needs to be lowered.

 { A similar case is a recent criticism in the results of ref.~\cite{Daniel} that I use as a constraint in Fig.~\ref{F1}. While ref.~\cite{Dutra2014,Rafa2011,LopesEPL} accept the results, using them as a constraint, others as ref.~\cite{PRLTan} suggest that those results are flawed, once they do not take into account finite temperature effects, as well the possibility of deconfinement. Nevertheless, the validity or not of the results from ref.~\cite{Daniel} will depend on additional HIC analyzes.}

{\bf Acknowledgments:} The author is eternally grateful to Debora P. Menezes for the many (MANY) English correction, and for almost seven years of tutelage.

\end{document}